# Scaling behavior of dynamic hysteresis in $Na_{0.5}Bi_{4.5}Ti_4O_{15}$ bulk ceramics

**Jianghao Huyan, Chunming Wang***

*School of Physics, Shandong University*

*No.27 Shandanan Road, Jinan, Shandong, China, 250100*

201400100120@mail.sdu.edu.cn, wangcm@sdu.edu.cn

## Abstract

The ferroelectric hysteresis loops of sodium bismuth titanate $Na_{0.5}Bi_{4.5}Ti_4O_{15}$ bulk ceramics were measured under periodical electric field in range of frequency from 0.01Hz to 100Hz and field from 10kV/cm to 150kV/cm. The three-stage scaling behavior of dynamic hysteresis was investigated in $Na_{0.5}Bi_{4.5}Ti_4O_{15}$ bulk ceramics. The scaling behavior at low amplitude of electric field is described as $\langle A\rangle \propto f^{0.122}E_0^{3.30}$ for low frequency and $\langle A\rangle \propto f^{0.122}E_0^{3.15}$ for high frequency. $\langle A\rangle$, $f$ and $E_0$ represent the area of hysteresis loop, frequency and amplitude of periodic electric field, respectively. At $E_0$ around coercive field, scaling behavior takes the form of $\langle A\rangle \propto f^{0.11}E_0^{4.28}$ for low frequency and $\langle A\rangle \propto f^{0.11}E_0^{4.17}$ for high frequency. At high $E_0$, we obtained $\langle A\rangle \propto f^{0.04}E_0^{2.90}$ for low frequency and $\langle A\rangle \propto f^{0.06}E_0^{2.75}$ for high frequency. The contribution to scaling relation mainly results from reversible of ferroelectric domain switching at low $E_0$, the velocity of domain wall motion at $E_0$ around coercive field and simultaneously reversible and irreversible domain switching at high $E_0$.

## 1. Introduction

Aurivillius bismuth layer-structured perovskites have drawn enormous attention on account of their rarely integrative properties including lead-free composition, electrically fatigue-free properties, excellent retention and relatively high Curie temperature Tc, leading them to be competitive candidate for several potential applications such as non-volatile ferroelectric random-access memory,[1-3] micro-electronic mechanical systems and high-temperature piezoelectric devices[4-6]. Hence a great many methods have been employed aiming to improve their piezoelectric properties aiming to promote their technical applications.[7-9] $Na_{0.5}Bi_{4.5}Ti_4O_{15}$(abbreviated as NBT) has layered structure which makes the improvements of NBT accessible and notable. However prior researches about NBT concentrated on its piezoelectric properties with little attention on ferroelectric properties.[9-11] The extensive applications of NBT is intrinsically related to the dynamics of domain reversal excited by the period-varying external electric field. Consequently it is of great significance to comprehensively study the ferroelectric domain reversal dynamics in NBT.

The area of hysteresis loop of ferroelectrics represents the energy dissipation during one cycle of the periodic external electric field. Furthermore the shape and area of the hysteresis loop have close relation with ferroelectric kinetics such as domain wall motion, domain nucleation and growth, domain rotation and domain structure.[12] In the past decades numerous theoretical and experimental studies have proposed the scaling law which is described as: $\langle A \rangle \propto f^{\alpha} E_0^{\beta}$ (where the exponent α and β depend on the dimensionality and symmetry of the sample). Rao et al. has studied the three-dimensional $(\Phi^2)^2$ and $(\Phi^2)^3$ models which is of O(N) symmetry with the large N limited, and the parameter in the scaling law can be theoretically determined as: $\langle A \rangle \propto f^{1/3} E_0^{2/3}$ at low frequency and $\langle A \rangle \propto f^{-1} E_0^{2}$ at high frequency.[13, 14] Barring the theoretical study, investigations of more materials have been reported and their results are various.[15-34]

NBT bulk ceramics, as a kind of three-dimensional ferroelectrics, exhibit different switching dynamic influenced by velocity of domain motion and domain nucleation rate. Additionally, scaling behavior of NBT bulk differs from that of NBT thin film due to depolarization effect, as a results of different texture intrinsically, including grain boundaries, domain walls, space charges and immobile defects.[27] Considering the crucial value of NBT in practical use it is imperative to detailedly study the dynamic hysteresis and scaling behavior of NBT which has not been reported before. Hence we commence this research to perfect the understanding of bismuth layered perovskites.

## 2. Experimental methods

The $Na_{0.5}Bi_{4.5}Ti_4O_{15}$ bulk ceramics was obtained conventional solid-state reaction. Analytical grades $Na_2CO_3$(99.8%), $Bi_2O_3$(99.8%) and $TiO_2$(99.9%) were prepared as starting materials. After weighed in calculated proportion, all the chemicals were wet mixed with ethanol in polyethylene bottles and were milled with $ZrO_2$ balls for 12h. The milled mixture was dried and later calcined at 770℃. Then the powders were milled in polyethylene again for 12h and dried after being milled. And the dried powders were granulated with polyvinyl alcohol (PVA) binder and then pressed into pellets at a pressure of 150Mpa. The NBT pellets were sintered at 1100℃ for 3h embedded in the NBT powders which is of same chemical composition with NBT pellets. The sintered pellets were polished up to 0.18mm thick and printed by sliver electrodes on both surfaces followed by baking at 550℃ for 30min. The ferroelectric hysteresis loops were obtained from aixACCT TF Analyzer 2000 (aixACCT CO., Germany). The electric field chosen varied from 10kV/cm-150kV/cm, which was limited by the breakdown voltage of NBT samples. The frequency range covered 0.05Hz-100Hz, and in details the range of frequency of a certain electric field was determined by the configuration of the equipment we use. Considering aixACCT TF Analyzer 2000 is limited to output maximum 2mA at high voltage unit, only the measurements, of which corresponding current is lower than 2mA, provide a high accuracy.

## 3. Results and discussion

The ferroelectric hysteresis loops at various external electric fields $E_0$ ($E_0$=10-150kV/cm) but fixed frequency $f$=2Hz are shown in Fig. 1a. It is obvious that in the low electric fields, the loop simply exhibit as slim loops initially. With the applied electric field increasing, the hysteresis loops at relatively high electric field start to saturate accompanied by the increasing $E_C$ $P_r$ and $\langle A \rangle$. Since $\langle A \rangle$ represents the energy dissipation during one cycle of the periodic external electric field, the increasing $\langle A \rangle$ with increasing $E_0$ indicates more energy is dissipated during the evolution of polarization driven by high electric field than low electric field. In order to investigate the dependence between $E_0$ and <A>, Fig. 1b reveals the three-stage linear relation between ln<A>and $lnE_0$. The range of applied electric field has been divided into three regions according to Fig. 1b: $E_0 \leq 70$kV/cm, 70kV/cm$< E_0 \leq$120kV/cm and 120kV/cm$< E_0 \leq$150kV/cm, which can be described as under coercive field (low electric field), near coercive field and above coercive field(high electric field). This three-stage dependence inspires us that the scaling relations of NBT at three regions are different.

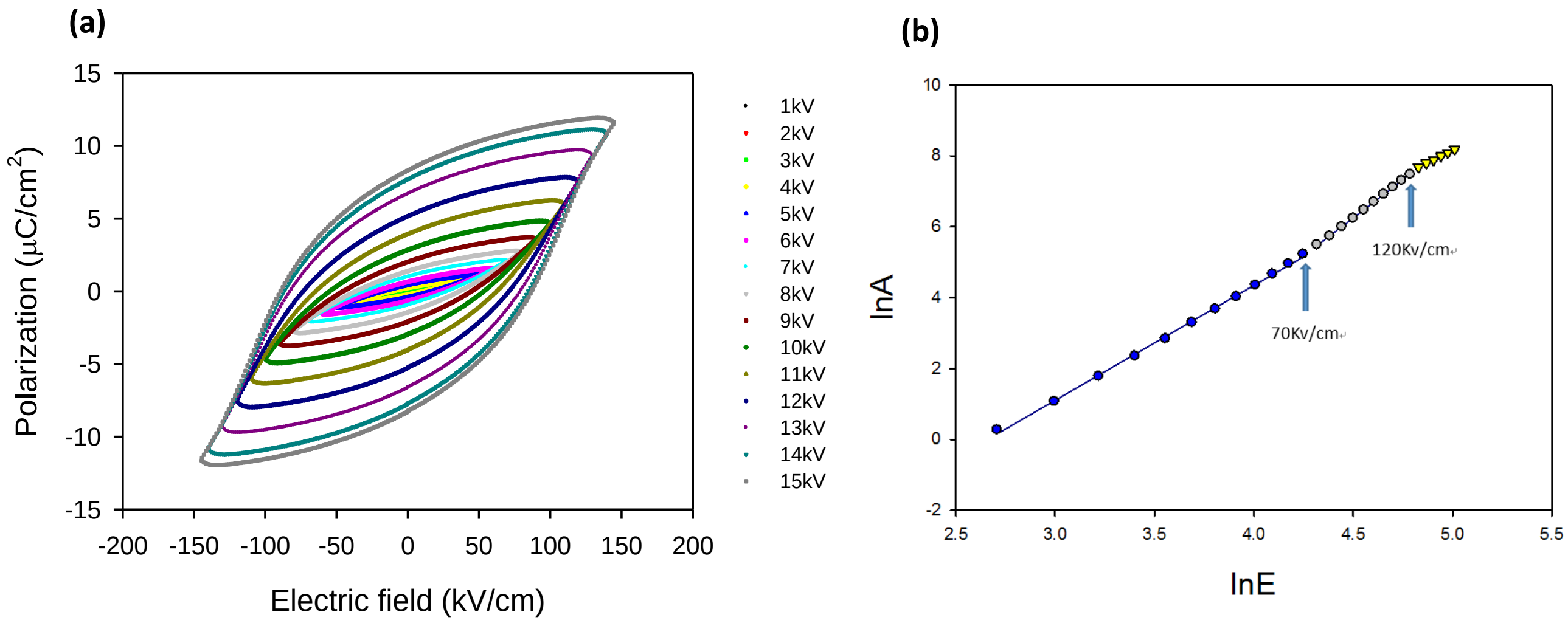

**Fig. 1 (a)** Hysteresis loops at various $E_0$; **(b)** relations of loop area <A> on $E_0$ (10kV/cm$\leq E_0 \leq$150kV/cm) at $f$=2Hz.

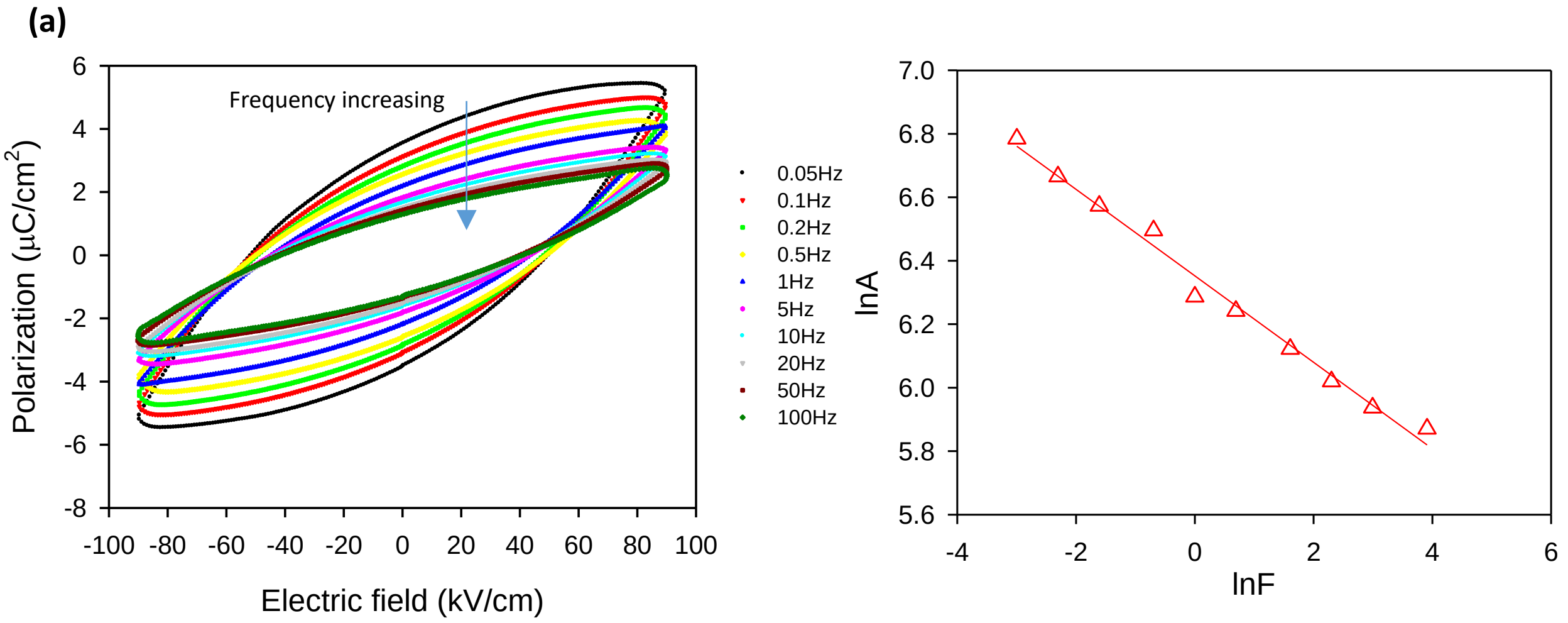

**Fig. 2 (a)** Hysteresis loops at various $f$ and $E_0$=90kV/cm; **(b)** relations of loop area <A> on $f$.

Fig. 2a shows the ferroelectric hysteresis loops measured at various frequencies $f$ ($f$=0.05-100Hz) but fixed applied electric field $E_0$=90kV/cm. The relation between frequency $f$ and area of the loop<A>is obtained shown by Fig. 2b. According to Fig. 2a, $P_r$ decreases and $E_C$ increases with the increasing frequency $f$. While the frequency of external electric field increasing, ferroelectric domains which respond to the changing electric field is decreasing. Namely increasing frequency means the period of the electric field becomes shorter, thus some ferroelectric domain switching cannot match up with the change of frequency, resulting the decline of $P_r$. Additional the $E_C$ can be considered as an effective field to overcome combination of the restoring force and viscous force during the domain wall motion.[35] At lower $f$, the velocity of domain wall motion is lower correspondingly therefore viscous force between domain walls could be neglected and $E_C$ becomes small. While f increasing, the velocity of domain wall motion increases correspondingly leading the considerable viscosity between domain walls. Consequently $E_C$ is increasing accompanied increasing amount of domain switching. Fig. 2b reveals the good linear relation of ln*f* and ln<A> and indicates a negative correlation of <A>and *f*.

To summarize the discussion above, NBT has different scaling behavior at different range of external electric field, which is divided into three regions by 70kV/cm and 120kV/cm. Furthermore the area of hysteresis loops<A>has positive correlation with applied electric field $E_0$ and negative correlation with frequency *f*. To comprehensively investigate the scaling behavior of NBT bulk ceramic, the dependence can be fitted into $\langle A\rangle \propto f^{a}E_0^{b}$, where the exponents a and b is directly obtained from experimental data.[27] By plotting <A> against f fixed $E_0$ and plotting <A> against $E_0$ fixed f, the exponents a and b are obtained respectively. The results are listed in the following:

| | | |
|---|---|---|
| $\langle A\rangle \propto f^{0.122}E_0^{3.30}$ | for low *f* and low $E_0$ | $R^2=0.984$ |
| $\langle A\rangle \propto f^{0.122}E_0^{3.15}$ | for high *f* and low $E_0$ | $R^2=0.993$ |
| $\langle A\rangle \propto f^{0.11}E_0^{4.28}$ | for low *f* and $E_0$ around $E_C$ | $R^2=0.981$ |
| $\langle A\rangle \propto f^{0.11}E_0^{4.17}$ | for high *f* and $E_0$ around $E_C$ | $R^2=0.998$ |
| $\langle A\rangle \propto f^{0.04}E_0^{2.90}$ | for low *f* and high $E_0$ | $R^2=0.981$ |
| $\langle A\rangle \propto f^{0.06}E_0^{2.75}$ | for high *f* and high $E_0$ | $R^2=0.996$ |

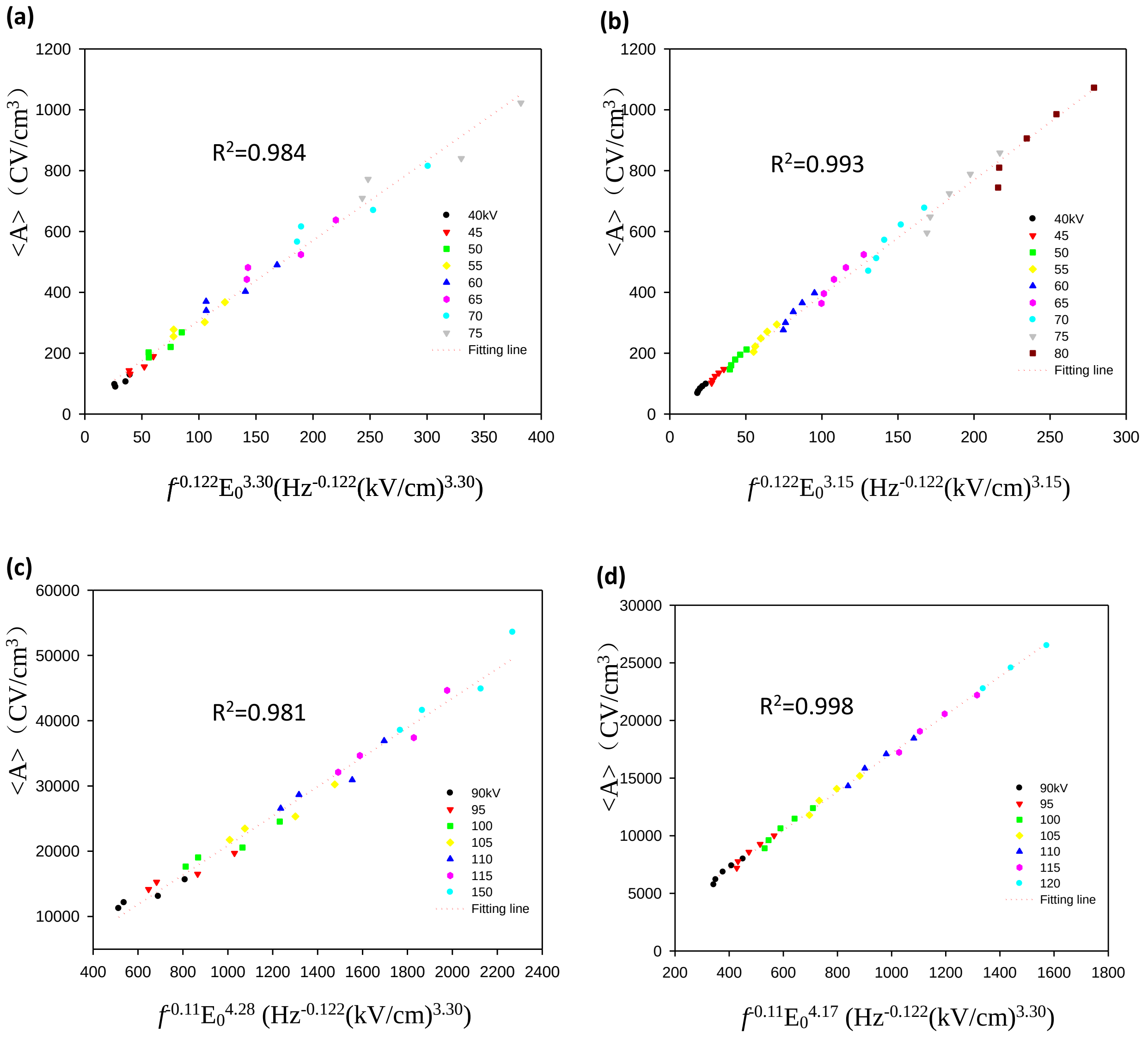

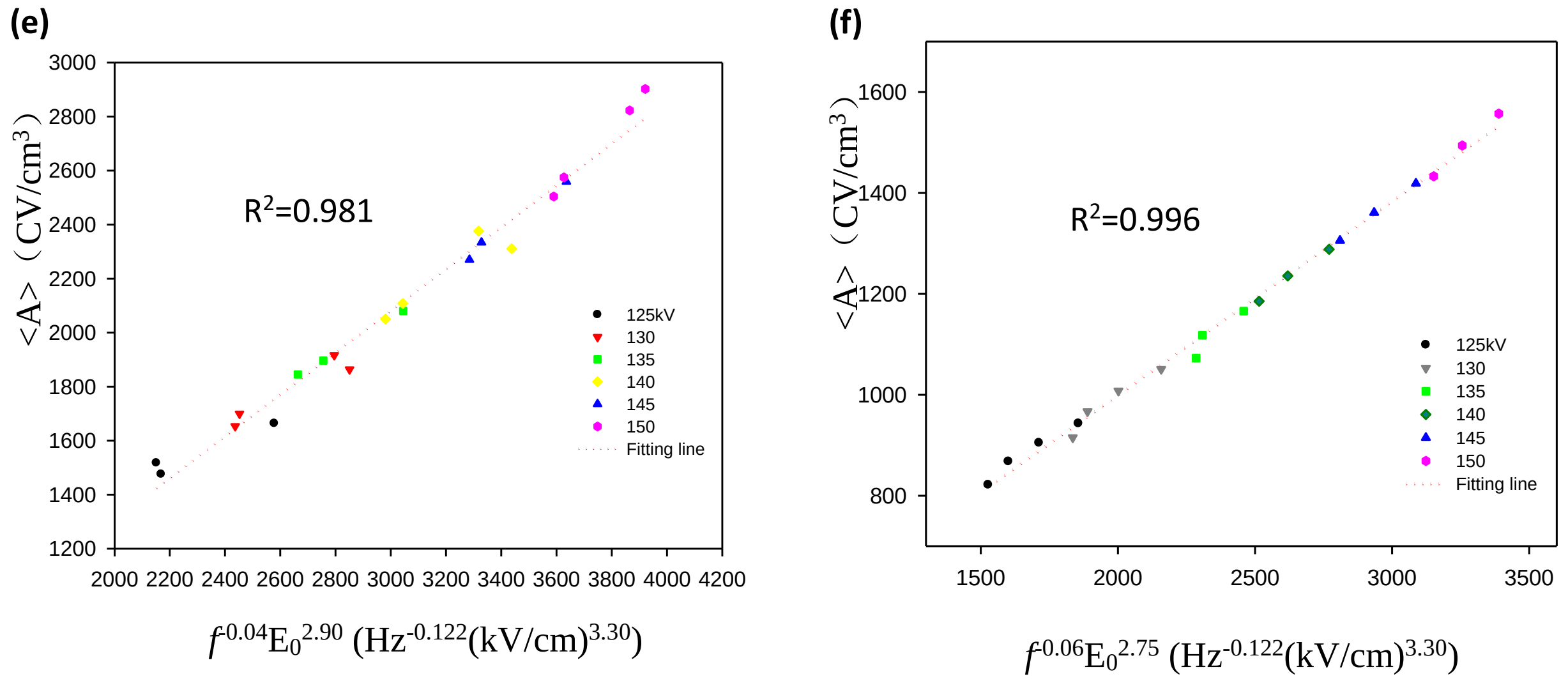

**Fig. 3** Scaling behavior of the loop area <A> against *f* and $E_0$ for NBT ceramics at room temperature (300K), **(a)** for low *f* and low $E_0$; **(b)** for high *f* and low $E_0$; **(c)** for low *f* and $E_0$ around $E_C$; **(d)** for high *f* and $E_0$ around $E_C$; **(e)** for low *f* and high $E_0$; **(f)** for high *f* and high $E_0$.

Some inferences could be obtained by analyzing the results of different $E_0$ regions. In the low $E_0$ region the exponent of *f* for low *f* equals to that for high *f* while the exponent of $E_0$ for low $E_0$ is larger than that for low $E_0$. The congruent negative exponent of *f* for low *f* and high *f* implies the domain reversal has the same negative correlation to frequency at low $E_0$ which can be explained by reversible polarization process. The ferroelectric hysteresis could be considered the results of reversible and irreversible polarization process which respond to the applied periodic-oscillating electric field.[36, 37] At low $E_0$ namely under the coercive electric field, the domain wall moves only in the minimum of random potential and only the reversible domain exists in the ferroelectrics. Hence most of the domain switching is reversible at low $E_0$ region leading the same exponent of frequency. The exponent of $E_0$ has a larger value at low *f* than high *f*. That is, energy dissipation is increasing more quickly under low-frequency electric field than high-frequency electric field. It is easy to understand that there is more time for domain to respond the change of low-frequency electric field than high-frequency electric field. Consequently more domain switching is reversible in the NBT driven by low-frequency electric field and the exponent for low *f* is larger than that for high *f*. In the region where $E_0$ is around $E_C$, the exponent for low *f* is larger than that for high *f*, which can be ascribed to the different amount of reversible domain switching. And the exponent for low *f* is same to that for high *f*, indicating that NBT has the same dependence on frequency at low *f* and high *f* under the electric field $E_0$ around $E_C$. In the high $E_0$ region, the exponent for frequency behaves differently at low *f* and high *f*, scaling behavior at low *f* has larger exponent for *f* than at high *f*. The absolute value of exponent for frequency at high *f* is larger than that at low *f*, namely the energy dissipation declines more quickly at high *f* than at low *f*, therefore NBT exhibits more sensitive under a high-frequency high electric field than a low-frequency high electric field. In other word, the increasing frequency leads the domain reversal to decline due to the flip time to short for domain to catch up with.

Considering the $E_0$ exponents integrally, more conclusions about evolution of polarization and domain structure could be drawn. The value of exponent for $E_0$ is the largest at $E_0$ around coercive electric field $E_C$, mainly due to the highest speed of domain wall motion and evolution of polarization in NBT. Driven by the external electric field around coercive field $E_C$, during the process that $E_0$ is initially under $E_C$ and is increasing up to exceed $E_C$, domains in NBT is applied enough high electric field to overcome the resistance caused by spontaneous polarization. Additionally, with the increasing $E_0$, the speed of domain wall motion is increasing correspondingly and the polarization in NBT has not totally saturated which offers domain walls great space to expand. Compared to the low $E_0$ region and high $E_0$ region, evolution of polarization is limited by resistance from coercive electric field and saturation due to volume limitation. Consequently the speed of polarization expansion is highest under $E_0$ around $E_C$, so the energy dissipation during a period of electric field is naturally increasing most quickly driven by $E_0$ around $E_C$. Fig. 3c and 3d show that the exponent for $E_0$ is larger at low $f$ than at high $f$. This is attributed to the reversible and irreversible domain switching process. With the frequency of electric field increasing, flip time of external electric field is getting shorter and shorter, thus more and more domain switching cannot respond to the change of electric field, namely amount of irreversible domain is increasing. Therefore energy dissipation increases faster at low $f$ than at high $f$, and exponent for $E_0$ is larger at low $f$ than at high $f$.

Considering the $f$ exponents integrally, the exponents for frequency at low $f$ is same to that at high $f$ under both low $E_0$ and $E_0$ around $E_C$. This indicates that when NBT is driven by low $E_0$ with both low frequency and high frequency, the domain switching is mostly reversible. On the other hand, the exponent for frequency at low $f$ is different to that at high $f$ under high $E_0$. This difference is caused by irreversible domain switching. That is, NBT driven by high electric field has different domain structure compared to NBT driven by low electric field. Three types of domain structures exist in bismuth-layered perovskites: the 180° domain, the 90° domain and the antiphase boundary.[38] Different types of domain are activated driven by different electric field. The 90° domain is activated at higher electric field than 180° domain, because the 90° domain switching is often accompanied with mechanical strain.[27] Thus the most contribution to domain switching in NBT driven by low $E_0$ is 180° domain reversal, while domain switching in NBT driven by high $E_0$ is combination of 180° domain and non-180° domain.

By analyzing the domain structure of NBT under high $E_0$ and low $E_0$, the different scaling dependence on $E_0$ is readily explainable. Due to different domain structure, the exponent for $E_0$ under low electric field is larger than under high electric field. In details, domain switching driven by low $E_0$ is mostly reversible whose domain structure exhibits 180° domain mostly, while domain switching driven by high $E_0$ has domain structure combined by 180° domain and non-180° domain. Hence the scaling behavior of dynamic hysteresis $E_0$ is not less sensitive to electric field at high than at low $E_0$. Thus scaling behaviors in high electric field differ from that in low electric field, and energy dissipation increases less quickly in high electric field owing to different contributions to domain structure in NBT bulk ceramics.

## 4. Conclusions:

The scaling behavior in $Na_{0.5}Bi_{4.5}Ti_4O_{15}$ was investigated and regression of experimental data revealed a three-stage relation in $E_0$ term. Scaling behavior of dynamic hysteresis can be described as:

$\langle A\rangle \propto f^{0.122}E_0^{3.30}$ for low $f$ and low $E_0$

$\langle A \rangle \propto f^{0.122}E_0^{3.15}$ for high $f$ and low $E_0$

$\langle A \rangle \propto f^{0.11}E_0^{4.28}$ for low $f$ and $E_0$ around $E_C$

$\langle A \rangle \propto f^{0.11}E_0^{4.17}$ for high $f$ and $E_0$ around $E_C$

$\langle A \rangle \propto f^{0.04}E_0^{2.90}$ for low $f$ and high $E_0$

$\langle A \rangle \propto f^{0.06}E_0^{2.75}$ for high $f$ and high $E_0$

The contribution to scaling relation mainly results from reversible of ferroelectric domain switching at low $E_0$, the velocity of domain wall motion at $E_0$ around coercive field and simultaneously reversible and irreversible domain switching at high $E_0$. Furthermore, at relatively high frequency the depolarization effect is a crucial factor which influences the frequency exponents in ferroelectric bulk ceramics.